\documentclass{article}

\usepackage{PRIMEarxiv}

\usepackage[utf8]{inputenc} % allow utf-8 input
\usepackage[T1]{fontenc}    % use 8-bit T1 fonts
\usepackage{hyperref}       % hyperlinks
\usepackage{url}            % simple URL typesetting
\usepackage{booktabs}       % professional-quality tables
\usepackage{amsfonts}       % blackboard math symbols
\usepackage{nicefrac}       % compact symbols for 1/2, etc.
\usepackage{microtype}      % microtypography
\usepackage{lipsum}
\usepackage{fancyhdr}       % header
\usepackage{graphicx}       % graphics
\graphicspath{{media/}}     % organize your images and other figures under media/ folder

\usepackage{latexsym}
\usepackage{amssymb}
\usepackage{amsmath}
\usepackage{amsthm}
\usepackage{booktabs}
\usepackage{enumitem}
\usepackage{graphicx}
\usepackage{color}
\usepackage{hyperref}
\usepackage{cleveref}

\title{Inequality in Congestion Games with Learning Agents}

\author{
  Dimitris Michailidis, Sennay Ghebreab, Fernando P. Santos \\
  Socially Intelligent Artificial Systems\\
  University of Amsterdam \\
  \texttt{\{d.michailidis, s.ghebreab, f.p.santos\}@uva.nl} \\
}

\begin{document}
\maketitle

\begin{abstract}
Who benefits from expanding transport networks? While designed to improve mobility, such interventions can also create inequality. In this paper, we show that disparities arise not only from the structure of the network itself but also from differences in how commuters adapt to it. We model commuters as reinforcement learning agents who adapt their travel choices at different learning rates, reflecting unequal access to resources and information. To capture potential efficiency-fairness tradeoffs, we introduce the Price of Learning (PoL), a measure of inefficiency during learning. We analyze  both a stylized network --- inspired in the well-known Braess's paradox, yet with two-source nodes --- and an abstraction of a real-world metro system (Amsterdam). Our simulations show that network expansions can simultaneously increase efficiency and amplify inequality, especially when faster learners disproportionately benefit from new routes before others adapt. These results highlight that transport policies must account not only for equilibrium outcomes but also for the heterogeneous ways commuters adapt, since both shape the balance between efficiency and fairness.
\end{abstract}

% keywords can be removed

\section{Introduction}
Expanding transport networks is often seen as an unambiguous good: more roads, more transport lines, more options means better access and faster travel. Yet, in practice, such interventions can backfire. Sometimes, adding infrastructure slows traffic---as showcased in Braess's Paradox \cite{braess_paradox}---and other times, it disproportionately benefits already advantaged areas, widening inequalities \cite{martens2016transport}.

Consider a city rolling out a new public transportation line intended to reduce crowdedness. While overall efficiency may improve, the gains may be unevenly distributed. Commuters with more resources or flexibility (say, access to better information, time to adapt, or higher risk tolerance), can adapt quickly. Others may not. The result? A transport upgrade that deepens divides.

These challenges expose a critical blind spot in how we model and evaluate urban mobility. Classical game theoretic models assume homogeneous, rational agents with perfect information. But real commuters behave differently: they experiment and adapt. And although research that investigates this type of adaptation are increasing \cite{carissimo_counter_intuitive,selten_commuters_2007,ben-elia_which_2010}, they have not yet been analyzed through the lens of fairness, especially when agents differ in their capacity to learn. In this paper, we use multi-agent reinforcement learning on congestion games to investigate how  exploration-driven learning shapes the distributional outcomes of transport expansions.

%Recent studies suggest that incorporating Reinforcement Learning (RL) dynamics can better capture this behaviour \cite{carissimo_counter_intuitive,selten_commuters_2007,ben-elia_which_2010}. 
% However, such adaptive processes have not yet been analyzed through the lens of fairness, especially when agents differ in their capacity to learn. In this paper, we use multi-agent reinforcement learning on congestion games to investigate how  exploration-driven learning shapes the distributional outcomes of transport expansions. %We build on the idea that differences in exploration behaviour and risk tolerance are correlated with socioeconomic factors like income, and can lead to unequal benefits from infrastructure changes.

A well-known example of unintended outcomes is the Braess's Paradox, which shows that adding a new route can paradoxically worsen congestion. Initially a  theoretical result, the  paradox has since been observed in real-world cities like New York, Boston, and Seoul \cite{youn2008price,baker2009removing}. Such paradoxes illustrate the power of game-theoretic models, which capture mobility as a strategic interaction among selfish agents \cite{roughgarden_severity_2006,Zhang_coordinated_route_2023}.

Yet these models rest on the assumption of perfectly rational, fully informed commuters. In practice, travellers adapt through trial and error: experimenting with routes, observing outcomes, and adjusting their choices over time. This process closely resembles Reinforcement Learning (RL), in which agents balance exploration (trying new strategies) with exploitation (relying on known ones) \cite{sutton_reinforcement_2018}. Empirical studies have shown that even simple RL mechanisms can successfully reproduce observed patterns in commuter behaviour \cite{selten_commuters_2007,ben-elia_which_2010}. Furthermore, modeling commuter behaviour through learning dynamics has shown deviations from the expected outcomes of full-information strategic agents --- agents do not necessarily converge to the Nash Equilibrium \cite{wunder2010classes,carissimo_counter_intuitive}. As such, introducing learning dynamics into traffic models not only increases realism but can also reshape expectations about system performance. Moreover, it enables the analysis throughout the entire adaptive process, rather than restricting attention solely to equilibria. We extend this line of research by tackling the fundamental question: \textbf{How do differences in learning affect fairness and efficiency in congestion games?}

Our question is further motivated by empirical research suggesting that wealthier individuals showcase a higher risk tolerance, and are generally more exploratory in their behaviours \cite{dong2020segregated, falleur_income_behavior}. For instance, \cite{dong2020segregated} find that residents of low-income neighborhoods are significantly less exploratory in their activities, effectively living within constrained environments. Psychological studies indicate that individuals from high-resource background demonstrate a higher tolerance for uncertainty and risk \cite{falleur_income_behavior}. These findings imply that exploration and learning rates can act as a proxies for resource capacity: individuals with greater resources are better equipped to search for and evaluate alternatives, while those with fewer resources are constrained by limitations on time, mobility, and finances. Such heterogeneity can exacerbate inequalities, yet it remains underexplored.

To address this gap, we introduce a framework that combines reinforcement learning dynamics with fairness analysis in congestion games. First, we propose the Price of Learning (PoL), a dynamic analogue of the Price of Anarchy that quantifies inefficiency during the learning process. Second, we analyse a stylized two-source network---informally referred to as Braess’ network, inspired by the classical Braess Paradox--- as a proof of concept to demonstrate how heterogeneous learning rates can amplify disparities. Third, we move to a real-world scenario and study an abstraction of the Amsterdam metro network, enabling us to examine commuter adaptation under realistic network expansions. Across both environments, our experiments show that interventions can worsen inequality: faster-learning groups capture greater benefits, while slower learners remain disadvantaged, even when efficiency is preserved. These results highlight the need for modelling to move beyond static analysis and consider how learning dynamics and behavioural heterogeneity shape long-run societal outcomes. To support reproducibility, we release our code \footnote{Github: \url{https://github.com/dimichai/fairness-braess}}.

\section{Related Work}
Our work relates to two themes in congestion games: learning dynamics within these games and the fairness of outcomes.

\subsection{Reinforcement Learning in Congestion Games}
Deviations from the standard rational commuter behaviour have been studied under models such as prospect theory, which consider uncertainty in travelers' decision-making \cite{chremos2022design,chremos_traveler-centric_2023}. Recent papers revealed non-Nash convergence and even chaotic behaviour. For instance, online learning algorithms such as Multiplicative Weights Update \cite{chotibut2020route} and Follow-the-Regularized-Leader \cite{bielawski2021follow} demonstrate that deviations from Nash equilibrium occur when considering learning agents. Learning automata have also been used  \cite{de_o_ramos_improved_2015}.

Reinforcement Learning is increasingly used to model travel choice behaviour in multiple congestion games \cite{ahmad2024travel}. Examples include studies on how commuters adapt over time through learning dynamics \cite{cats2020learning,ben-elia_which_2010}, the impact of memory \cite{wei_day_to_day}, as well as the impact of tolls in congestion games \cite{ramos2020toll}.

Analyzing the effects of learning in congestion games through Reinforcement Learning, and in particular, multi-agent Q-learning hasn't been explored thoroughly. Some efforts have achieved this in a parallel road network, showing that the total social cost varies with different exploration rates \cite{levy2018emergence}. In particular, they show that with sufficiently small exploration, the system will converge to the inefficient equilibrium. Others investigate the effect of learning on a system’s convergence between Nash equilibrium and the social optimum \cite{carissimo_counter_intuitive}. However, while they assume a homogeneous population with a uniform learning rates, in reality, commuters can show heterogeneity in learning. Here, we build upon this work and extend this by modeling two distinct populations with varying learning rates, allowing for a richer analysis of strategic behaviour and its implications for equilibrium outcomes, on both a stylilzed and a real-world network.

\subsection{Fairness in Congestion Games}

The challenge of avoiding increased costs after network expansions, as showcased by the well-known Braess’s Paradox, has been extensively studied assuming selfish agents, with efforts focused on network design \cite{roughgarden_severity_2006,gairing_network_design_2017,roman2019multi}.

Our work contributes to this literature by examining fairness considerations in congestion games, an aspect previously explored primarily in centrally coordinated settings. Prior research has explored fairness through various mechanisms. One line of work analyzed the use of taxation to balance fairness and efficiency \cite{ferguson2021impact}, while another investigated artificial currency–based pricing schemes as a means to promote equitable outcomes \cite{pedroso2024fair}. A karma-based framework for achieving fairness was proposed in related studies \cite{censi2019today}. Further, the Double Braess’s Paradox was introduced in a multi-resource setting with two sources and two destinations, where action-restriction mechanisms were evaluated as a fairness intervention \cite{oesterle2024raise}. Additional work on weighted congestion games has examined mechanisms designed to reduce disparities in agent outcomes \cite{fischer2023fair}.

Using agent-based models, prior work has examined how individual behaviors affect fairness and efficiency in network settings. One study analyzed a five-edge Braess’s paradox network, evaluating the impact of various micro-level parameters on system performance \cite{Belov04072022}. Another line of research investigated fairness in travel times between two routes through the lens of learning dynamics, assuming homogeneous exploration and learning rates \cite{levy2018emergence}.

Our approach examines fairness as an emergent property in a multi-agent system. We define fairness as the disparity in average travel time to the destination across different sources, influenced by varying learning rates. Our model assumes learning agents who adapt their behaviour through reinforcement learning.

Unlike previous studies, our work explicitly models two distinct populations with varying learning capacities. This distinction enables a more nuanced analysis of fairness, shedding light on its interaction with strategic behaviour and resource access.
\section{Preliminaries}
\label{subsec:preliminaries}
% We define congestion games, Nash equilibrium, price of anarchy, price of learning, and fairness.

\subsection{Congestion Games}
We consider standard congestion games \cite{roughgarden_severity_2006}. Let $G = (V, E)$ be a directed graph, where $V$ is the set of nodes and $E$ the set of edges. Let $\mathcal{N}$ be the set of players with $N = |\mathcal{N}|$, where each player $i \in \mathcal{N}$ has a source node $s_i \in V$ and a common destination $d \in V$. 

Each edge $e \in E$ is associated with a latency function
\begin{equation}
f_e : \mathbb{R}_{\ge 0} \times \mathbb{R}_{>0} \to \mathbb{R}_{\ge 0}, \quad (x,K_e)\mapsto f_e(x;K_e),
\end{equation}

where $x$ is the edge load and $K_e > 0$ is the capacity. We allow for two edge types:
\begin{itemize}
    \item \textbf{Flow-sensitive edges:} latency depends on both $x$ and $K_e$.
    \item \textbf{Fast or constant-cost edges:} latency is constant and does not depend on capacity.
\end{itemize}
Formally, this can be written as
\[
f_e(x;K_e) =
\begin{cases} 
f_e^\text{flow}(x;K_e), & \text{if edge $e$ is flow-sensitive},\\
f_e^\text{const}, & \text{if edge $e$ is fast/constant}.
\end{cases}
\]

We assume:

\begin{itemize}
    \item \textbf{Nonnegativity:} $f_e(x) \ge 0$ for all $x \in \mathbb{N}$.
    \item \textbf{Continuity and Monotonicity:} $f_e$ is continuous and nondecreasing \cite{roughgarden2002unfair}.
    \item \textbf{Polynomial-time computation:} $f_e(x)$ can be computed in polynomial time \cite{fischer2023fair}.
\end{itemize}

Each player $i$ selects a strategy $a_i \in \mathcal{A}_i$, where $\mathcal{A}_i$ is the set of all paths from $s_i$ to $d$. The set of all strategy profiles is $\mathcal{A} = \prod_{i \in \mathcal{N}} \mathcal{A}_i$, and a profile is denoted $a = (a_1, \dots, a_N) \in \mathcal{A}$.  

The edge load is:
\begin{equation}
x_e(a) = |\{ i \in \mathcal{N} : e \in a_i \}|,
\end{equation}
and the individual cost of player $i$ is:
\begin{equation}
c_i(a)=\sum_{e\in a_i} f_e\big(x_e(a);K_e\big),
\end{equation}

The social cost is the sum of all individual costs:
\begin{equation}
C(a)=\sum_{e\in E} x_e(a)\, f_e\big(x_e(a);K_e\big).
\end{equation}

\subsection{Nash Equilibrium}
A strategy profile $a^{\text{NE}} \in \mathcal{A}$ is a \textit{Nash equilibrium (NE)} if no player can reduce their individual cost by unilaterally deviating:
\begin{equation}
\forall i \in \mathcal{N}, \quad \forall a_i' \in \mathcal{A}_i: \quad c_i(a_i^{\text{NE}}, a_{-i}^{\text{NE}}) \le c_i(a_i', a_{-i}^{\text{NE}}).
\end{equation}

Since congestion games are potential games \cite{monderer_potential_1996, milchtaich1996congestion}, at least one pure-strategy NE exists \cite{rosenthal1973network}.

\subsection{Social Optimum, Price of Anarchy, and Price of Learning}
The \textit{Social Optimum (SO)} is the strategy profile that minimizes total cost, and is defined as
\begin{equation}
\label{eq:chapter5:so}
a^\star \in \arg \min_{a \in \mathcal{A}} C(a).
\end{equation}

The \textit{Price of Anarchy (PoA)} measures inefficiency of NE relative to the social optimum \cite{roughgarden_2002_how_bad_selfish_routing}:
\begin{equation}
\label{eq:chapter5:poa}
\text{PoA} = \frac{\max_{a \in \text{NE}} C(a)}{C(a^\star)} \ge 1.
\end{equation}

While the PoA evaluates inefficiency under rational, fully informed behaviour, in many practical settings players must learn their strategies over time. In such cases, their behaviour is better modeled through adaptive processes, rather than immediate convergence to NE. To analyze this setting, we extend the notion of inefficiency to account for possibly non-equilibrium policies, by introducing the \textit{Price of Learning}.

We define a \textit{policy} as the function that specifies an agent’s behavior in response to its observed state. Formally, the policy of agent $i$ at time $t$ is a mapping from a state $\mathcal{S}^i_t$ to an action in $\mathcal{A}$. Let $\pi^i_t$ denote the policy of agent $i$ at time $t$, and let $\pi_t = (\pi^1_t, \ldots, \pi^n_t)$ denote the \textit{joint policy} (a set of individual policies) at time $t$. Let $A_t \sim \pi_t$ be a random strategy profile drawn from this joint policy. The expected social cost under $\pi_t$ is
\begin{equation}
\label{eq:chapter5:pol}
C(\pi_t) = \mathbb{E}_{A_t \sim \pi_t}[C(A_t)],
\end{equation}
and the \textit{Price of Learning (PoL)} at time $t$ is
\begin{equation}
\text{PoL}(t) = \frac{C(\pi_t)}{C(a^\star)} \ge 1,
\end{equation}
where $a^\star$ is the socially optimal profile (\Cref{eq:chapter5:so}).

The PoL quantifies the inefficiency of the learning process at a given time. It is a dynamic measure of system performance, that bridges the gap between equilibrium and learning-based dynamics.

\subsection{Fairness}
We assess fairness through the notion of source disparity, defined as the difference in average cost between groups of players originating from different sources. Our definition is inspired by fairness (or justice) concepts in urban planning, where location is central: geographic position determines accessibility to amenities and thus shapes opportunities \cite{martens2016transport}. Our metric requires that no source node is systematically disadvantaged in the costs incurred to reach a destination. From the perspective of artificial intelligence fairness, this corresponds to the group fairness criterion of independence, which demands that outcomes---here, travel costs---remain independent of an agent’s point of origin \cite{castelnovo_clarification_2022}.

Let $\mathcal{N}_j \subseteq \mathcal{N}$ denote the set of players starting from source $s_j$ with $N_j = |\mathcal{N}_j|$. The average cost for source $s_j$ is
\begin{equation}
\text{AvgCost}(s_j) = \frac{1}{N_j} \sum_{i \in \mathcal{N}_j} c_i(a), \quad j=1,2,
\end{equation}
and the source disparity is
\begin{equation}
\label{eq:chapter5:sd}
\text{SD}(s_1, s_2) = \text{AvgCost}(s_1) - \text{AvgCost}(s_2).
\end{equation}

A positive SD favors $s_2$, negative favors $s_1$, and zero indicates perfect fairness.

\section{Learning Dynamics}
In practice, humans deviate from exact Nash equilibrium strategies, as they lack complete information about which strategies yield specific payoffs \cite{bonau_begavioural_game_theory, lee_game_2008}. Instead of playing optimally from the outset, individuals tend to explore various options and gradually learn to favor those that maximize their satisfaction. This adaptive behavior has been studied in congestion games, where outcomes frequently diverge from Nash equilibrium predictions \cite{carissimo_counter_intuitive}. For this reason, we argue that it is essential to study the process itself, rather than assuming that commuters will ultimately converge. To achieve this, we use a framework in which agents repeatedly play the game and adapt their strategies through Reinforcement Learning (RL).

Importantly, we employ RL not as a tool for optimisation, but to simulate the adaptive behaviours of agents interacting within complex environments. Our perspective aligns with the \textit{descriptive} agenda in multi-agent learning, which emphasizes modeling learning processes that mirror how real-world agents, such as humans, adapt in social contexts \cite{SHOHAM2007365}. Importantly, this focus is not limited to human learning. The same framework extends to RL-based systems themselves. For example, as RL-powered recommender systems become increasingly prevalent \cite{afsar_rl_survey}, it is likely that real-time adaptive systems will also be deployed in navigation applications. Understanding their learning dynamics is therefore equally crucial.

Here, we adopt a Q-learning framework, a model-free reinforcement learning algorithm, to capture some simple properties, such as trial-and-error adaptation, and bounded rationality. This choice is grounded in that commuters may lack comprehensive or accurate models of how their actions influence congestion dynamics. Instead, they rely on repeated experience to guide future choices. Model-free learning allows us to study emergent behaviours under simple, reactive update rules, without assuming agents have detailed planning capabilities or perfect knowledge.

\subsection{Q-Learning Framework}
We model repeated play over $T$ discrete time steps, $t = 0,1,\dots,T-1$. Each player $i \in \mathcal{N}$ seeks to minimize experienced cost (travel time plus congestion) by updating estimates of expected cost for each strategy $a_i \in \mathcal{A}_i$.

\textbf{Q-values:} Each player maintains $Q_i(a_i)$, estimating cumulative cost for selecting $a_i$.  

\textbf{Update rule:}
\begin{equation}
Q_i(a_i) \gets Q_i(a_i) + \alpha \Big[r_i(t) + \gamma \max_{a_i' \in \mathcal{A}_i} Q_i(a_i') - Q_i(a_i)\Big],
\end{equation}
where $\alpha \in (0,1]$ is the learning rate, $\gamma \in [0,1]$ the discount factor, and the observed reward is
\begin{equation}
r_i(t) = -\sum_{e \in a_i(t)} f_e(x_e(a_t); K_e),
\end{equation}
with $a_i(t)$ the strategy chosen at time $t$ and $a_t$ the realized profile. The negative sign ensures Q-learning maximizes the negative latency (minimizes actual latency). We assume the game is stateless and rewards are observed immediately (as in, e.g., \cite{claus1998dynamics,carissimo_counter_intuitive}). Agents starting in different sources may have different learning rates.

\textbf{Action selection:} Each agent follows an $\epsilon$-greedy policy $\pi_t^i$:
\[
a_i(t) =
\begin{cases}
\text{randomly pick from } \mathcal{A}_i, & \text{with probability } \epsilon_i, \\
\arg\max_{a_i \in \mathcal{A}_i} Q_i(a_i), & \text{with probability } 1-\epsilon_i.
\end{cases}
\]

\textbf{Convergence}
At each step, agents collectively contribute to congestion. Each agent estimates the expected cost of selecting a given strategy. With continued learning, Q-values converge to estimates reflecting expected cost. However, these do not necessarily correspond to a Nash Equilibrium (NE); convergence is shaped by factors such as the learning rate, exploration rate, and the environment’s structural properties \cite{carissimo_counter_intuitive}. Unlike traditional RL aimed at maximizing a reward, our focus here is on describing and understanding emergent phenomena in a population of learning agents.

\begin{figure}[t!]
\centering
\includegraphics[width=2in]{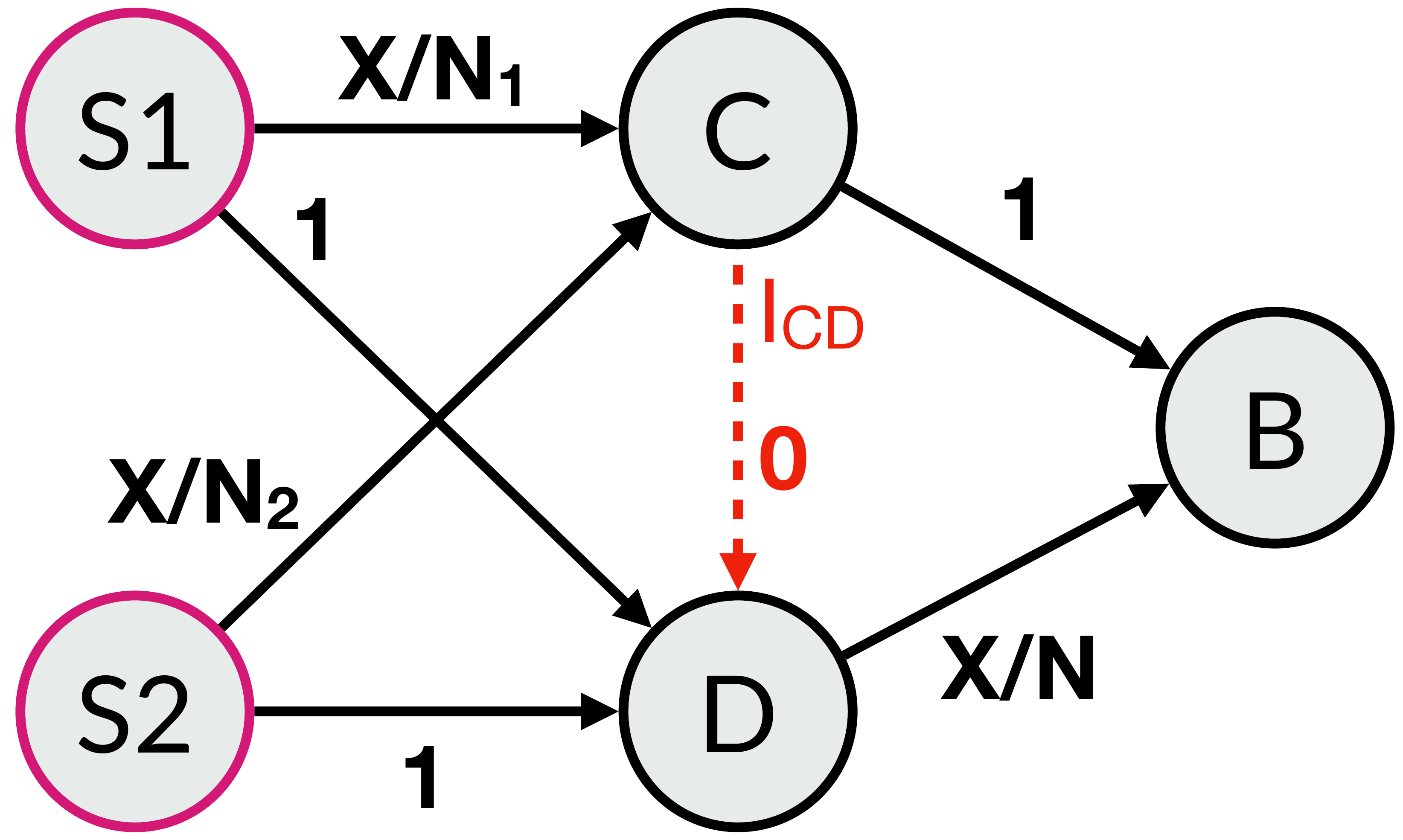}
\caption{A Braess's Paradox game with two sources ($S1$ \& $S2$) connecting to a single destination node $B$. We examine two phases: before and after the network extension $I_{CD}$ (red link).}
\label{fig:chapter5:game}
\vspace{-3mm}
\end{figure}

\section{Environments}
We introduce two routing games: a stylized two-source extension of Braess’s paradox (\Cref{subsec:chapter5:braess_game}) that illustrates how added capacity can worsen outcomes, and an abstraction of the Amsterdam metro network (\Cref{subsec:chapter5:amsterdam_metro}) to study commuter adaptation under different network phases. We conclude with a description of the experimental setup used to evaluate learning dynamics (\Cref{subsec:chapter5:experimental_setup}).

\begin{table}
\centering
\begin{tabular}{lcccc}
\hline
Two-Source Braess                & \textbf{NE}             & \textbf{SO}             & \textbf{PoA}            & \textbf{SD}              \\ \hline
(1) No Intervention              & 300                     & 300                     & 1.0                     & 0.0                      \\
(2) I$_{\text{CD}}$ Intervention & 400                     & 300                     & 1.33                    & 0.0                      \\ \hline
Amsterdam Metro                  & \textbf{NE}             & \textbf{SO}             & \textbf{PoA}            & \textbf{SD}              \\ \hline
(A) Before North-South & \multicolumn{1}{l}{9700} & \multicolumn{1}{l}{9700} & \multicolumn{1}{l}{1.0} & \multicolumn{1}{l}{27.00} \\
(B) After North-South            & \multicolumn{1}{l}{7349} & \multicolumn{1}{l}{7349} & \multicolumn{1}{l}{1.0} & \multicolumn{1}{l}{13.75} \\
(C) After West-Amstel            & \multicolumn{1}{l}{6648} & \multicolumn{1}{l}{6648} & \multicolumn{1}{l}{1.0} & \multicolumn{1}{l}{4.52} \\ \hline
\end{tabular}%
\caption{Overview of equilibrium results, without learning and $100$ agents starting at each source. We report total cost at the Nash Equilibrium (NE), Social Optimum (SO), Price of Anarchy (PoA, Eq. \ref{eq:chapter5:poa}), average Source Disparity (SD, Eq. \ref{eq:chapter5:sd}).}
\label{tab:chapter5:ne_so}
\end{table}

\begin{figure*}[t!]
    \centerline{\includegraphics[width=6in]{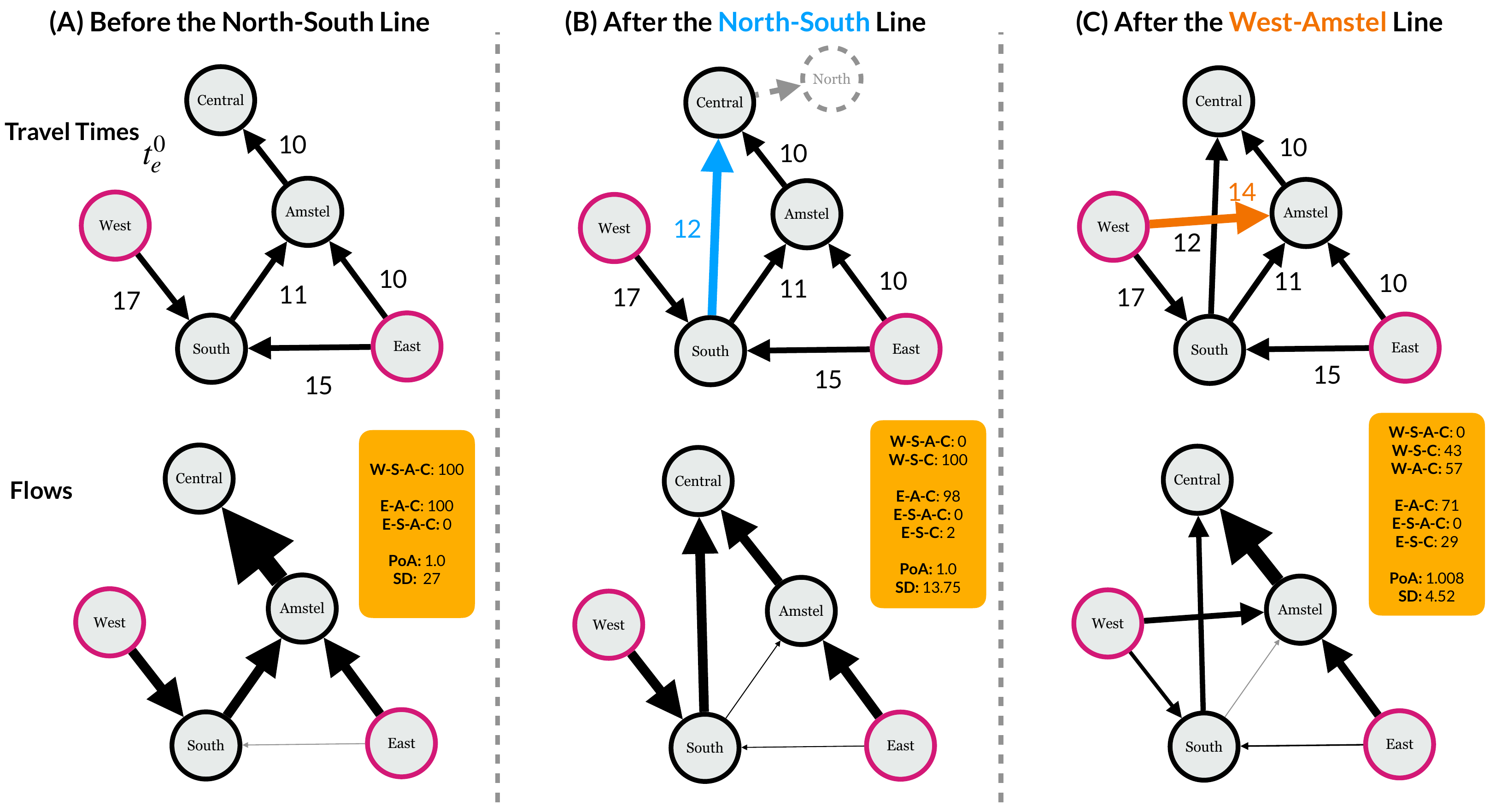}}
    \caption{Top: The Amsterdam Metro Environment with key stations and routes for two commuter sources (West and East) converging at Central (See the Appendix for an overlay on the actual Metro Network Map). Three network phases are shown: Phase A (pre-2018, no North–South line), Phase B (current, with North–South line), and Phase C (future expansion connecting West and East via Amstel). Edge labels indicate free-flow travel times. Bottom: Arrow sizes indicate flows under the Nash Equilibrium, illustrating route choices and congestion distribution.}
    \label{fig:chapter5:amsterdam_metro}
\end{figure*}

\subsection{A Two-Source Braess's Paradox Game}
\label{subsec:chapter5:braess_game}
The Braess’s paradox demonstrates that adding capacity to a network can paradoxically increase congestion \cite{braess_paradox}. Although often illustrated using stylized networks, it is neither rare nor anomalous under realistic assumptions \cite{steinberg1983prevalence}, and has been observed empirically in cities such as New York, Boston, and Seoul \cite{youn2008price,baker2009removing}.

To account for urban segregation effects and the observation that different demographic groups can systematically start commuting routes from different locations, we consider an extended Braess network with two source nodes and a single destination, illustrated in \Cref{fig:chapter5:game}. Two configurations are studied: (1) without the fast lane, where both sources have two strategies to reach the destination; and (2) with a zero-cost link from node $C$ to $D$, giving each source a third strategy.

Let $N_1 = |\mathcal{N}_1|$ and $N_2 = |\mathcal{N}_2|$ denote the number of players starting from $S_1$ and $S_2$, respectively. Each player selects a strategy $a_i \in \mathcal{A}_i$ corresponding to a path to $B$.

The latency functions $f_e(x;K_e)$ for edges $e \in E$ follow the general framework we defined in \Cref{subsec:preliminaries}. Specifically: for edges $S_1 \to C$ and $C \to B$, the latency is $f_e(x;K_e) = x / N_1$; for edges $S_2 \to D$ and $D \to B$, the latency is $f_e(x;K_e) = x / N_2$; for edges $S_1 \to D$ and $S_2 \to C$, the latency is constant: $f_e(x;K_e) = 1$. When the additional \textit{fast lane} $C \to D$ is introduced, it has constant latency $f_e(x;K_e) = 0$, mimicking the original Braess network.

The latency functions $f_e(x)$ for some edges $e \in E$ depend on the number of players $x$ using the edge and are defined as follows: for edges $S1 \to C$ and $C \to B$, the latency is $f_e(x) = x / N_1$; for edges $S2 \to D$ and $D \to B$, the latency is $f_e(x) = x / N_2$; for edges $S1 \to D$ and $S2 \to C$, the latency is a constant $f_e(x) = 1$. When the additional \textit{fast lane} $I_{CD}$ is introduced, it has a constant flow $f_e(x) = 0$, mimicking the original Braess network.

\textbf{Strategies:}

\textbf{Before intervention}: $a_i \in \{$Up: $C \to B$, Down: $D \to B\}$.

\textbf{After intervention}: $a_i \in \{$Up, Down, Cross: $C \to D \to B\}$.

% \textbf{Edge loads:} $x_e(a) = |\{ i \in \mathcal{N} : e \in a_i\}|$.  

\textbf{Social cost:}
\begin{equation}
\label{eq:chapter5:social_cost}
C(a) = \sum_{e \in E} x_e(a)\, f_e\big(x_e(a);K_e\big).
\end{equation}

The game has non-decreasing latency functions, and thus, by Rosenthal’s theorem \cite{rosenthal1973network}, admit at least one pure-strategy Nash equilibrium. We provide the detailed construction of the potential functions and full proof of equilibrium existence in the Appendix.

As shown in \Cref{tab:chapter5:ne_so}, in the baseline network, players from $S_1$ and $S_2$ converge to a socially optimal state. The intervention introduces a third strategy (Cross) for all players, shifting the Nash equilibrium so that all choose the new path, thereby increasing total social cost and the price of anarchy.

\subsection{The Amsterdam Metro Network Game}
\label{subsec:chapter5:amsterdam_metro}

To investigate how commuters adapt to changes in real-world transportation systems, we abstract the Amsterdam Metro Network into a routing game. Two origin groups of players start at $W$ and $E$, converging toward $C$ via intermediate hubs $S$ and $A$. Let $\mathcal{N}_W$ and $\mathcal{N}_E$ denote the player sets with $N_W$ and $N_E$ players, respectively, with $N = N_W + N_E$. The network structure is depicted in \Cref{fig:chapter5:amsterdam_metro} (top). See the Appendix (Figure 2) for details on how the real metro network is translated into the network we consider in this section. 

The model emphasizes the busiest stations in the Amsterdam metro network: Sloterdijk (denoted as \textbf{W}est), Zuid (\textbf{S}outh), Bijlmer Arena (\textbf{E}ast), \textbf{A}mstel, and Centraal (\textbf{C}entral), which is the city's primary commuter hub and is modeled as the unique destination. The game is examined under three distinct network phases. 

\textbf{Network phases:} Phase A (past): Pre-2018, without North–South line. Phase B (current): Includes North–South line. Phase C (hypothetical): Future expansion connecting \textbf{W}est and \textbf{E}ast via \textbf{A}mstel.

\textbf{Strategies:}

\textbf{Phase A}: West: $a_W \in \{W \to S \to A \to C\}$ East: $a_E \in \{E \to A \to C, E \to S \to A \to C\}$.

\textbf{Phase B}: West: $a_W \in \{W \to S \to A \to C, W \to S \to C\}$ East: $a_E \in \{E \to A \to C, E \to S \to A \to C, E \to S \to C\}$.

\textbf{Phase C}: West: $a_W \in \{W \to S \to A \to C, W \to S \to C, W \to A \to C\}$ East: $a_E \in \{E \to A \to C, E \to S \to A \to C, E \to S \to C\}$.

Initially, West has a single strategy (as in the Metro network pre-2018), expanding with each network phase. East starts with two strategies and reaches three by Phase C.

Player cost is modeled as a function of congestion using a volume–delay function widely applied in transport planning \cite{das_link_cost_function}.

\textbf{Social Cost:}  
For each edge $e$ with load $x_e$:
\begin{equation}
f_e(x_e;K_e) = t_e^0 \left( 1 + \frac{x_e}{K_e} \right),
\end{equation}

where $t_e^0$ is the free-flow travel time and $K_e$ is the edge capacity. Free-flow times are calculated from real-world schedules (details in the Appendix). This function is appropriate for public transport networks, where both travel time and crowding influence modal choice \cite{li_crowding_pt}. Moreover, empirical evidence from Amsterdam shows nearly half of commuters adjust behaviour in response to overcrowding \cite{monitor_sociale_2023}. For simplicity, we set a uniform capacity $K_e = N$ across all edges, implying that the cost of an edge doubles when it is fully occupied compared to when it is empty.

Consistent with the Braess paradox, the total system cost $C$ is the sum of individual link costs (Eq. \ref{eq:chapter5:social_cost}). \Cref{fig:chapter5:amsterdam_metro} (Top) shows the three network phases and their free-flow travel times $t_e^0$. Similarly to the Braess Network, the metro game admits at least one pure-strategy Nash equilibrium (see Appendix for detailed construction of the potential functions and full proof of equilibrium existence). \Cref{tab:chapter5:ne_so} reports NE, SO, PoA, and SD, assuming no learning agents. In \Cref{fig:chapter5:amsterdam_metro} (Bottom), we show the flows in the network under the NE. A key distinction between the Amsterdam and Braess networks is that Amsterdam does not exhibit a Price of Anarchy: NE and SO coincide in all phases, and selfish routing is efficient under classical game theory assumption.

Analyzing the NE further, network extensions consistently improve fairness and efficiency. The Source Disparity decreases, and PoA remains at $1$. With the North–South line, agents from West switch to the West $\to$ South $\to$ Central route, decongesting Central–Amstel, consistent with empirical observations in Amsterdam, where North-South line became the most used one after its opening \cite{gvb_north_south_line}. Phase C shifts congestion slightly back to Amstel–Central, but to a lesser extent than in Phase A. We explore whether learning agents converge to a similar equilibrium.

\subsection{Experimental Setup}
\label{subsec:chapter5:experimental_setup}

We conducted experiments with $|\mathcal{N}| = 200$ agents, evenly divided between the two sources in both games. Each agent’s Q-values were initialized uniformly at random. To ensure that observed effects can be attributed to the learning rate, we applied a common exploration strategy across all agents: the exploration rate $\epsilon$ was decayed exponentially (details provided in the Appendix). This procedure enabled agents to transition gradually from exploration to exploitation over the course of training. Each experiment was run for $10{,}000$ steps and repeated $40$ times with independent random seeds to ensure robustness of the results. We designed two sets of experiments to isolate the role of the learning rate: 

\textbf{Equal rates across sources.} Agents from both sources were assigned identical learning rates. We systematically varied the common learning rate across multiple values to examine how, under symmetric conditions, agents adapt to interventions. 

\textbf{Asymmetric rates across sources.} To study fairness implications arising from differences in learning, we fixed the \emph{average} learning rate across the population at $0.2$, while varying the  ratio of learning rates between the two sources. Specifically, we examined the set of ratios: $\{5/1, 4/1, 3/1, 2/1, 1/1, 1/2, 1/3, 1/4, 1/5\}$. For example, a $5/1$ ratio indicates that one source (S1 in Braess, W in Amsterdam) learns five times faster than the other (S2 and E).

In the results section, we indicate whether figures correspond to experiments with absolute learning rates or relative ratios.

% This Q-learning setup allows us to study the emergent behaviours in the two-source Braess's paradox game under repeated play. Specifically, it enables us to assess whether reinforcement learning dynamics amplify or mitigate the inefficiencies of selfish routing. Additionally, it allows us to assess the impact of disparities in exploration rates between the two sources on the price of anarchy and fairness.

\section{Results}
We first analyze learning dynamics on the Braess network, and then extend our study to the Amsterdam Metro network.

\subsection{Results on the Braess Network}

\subsubsection{Unequal learning rates amplify disparities}
\begin{figure}[t!]
\centering
\includegraphics[width=4in]{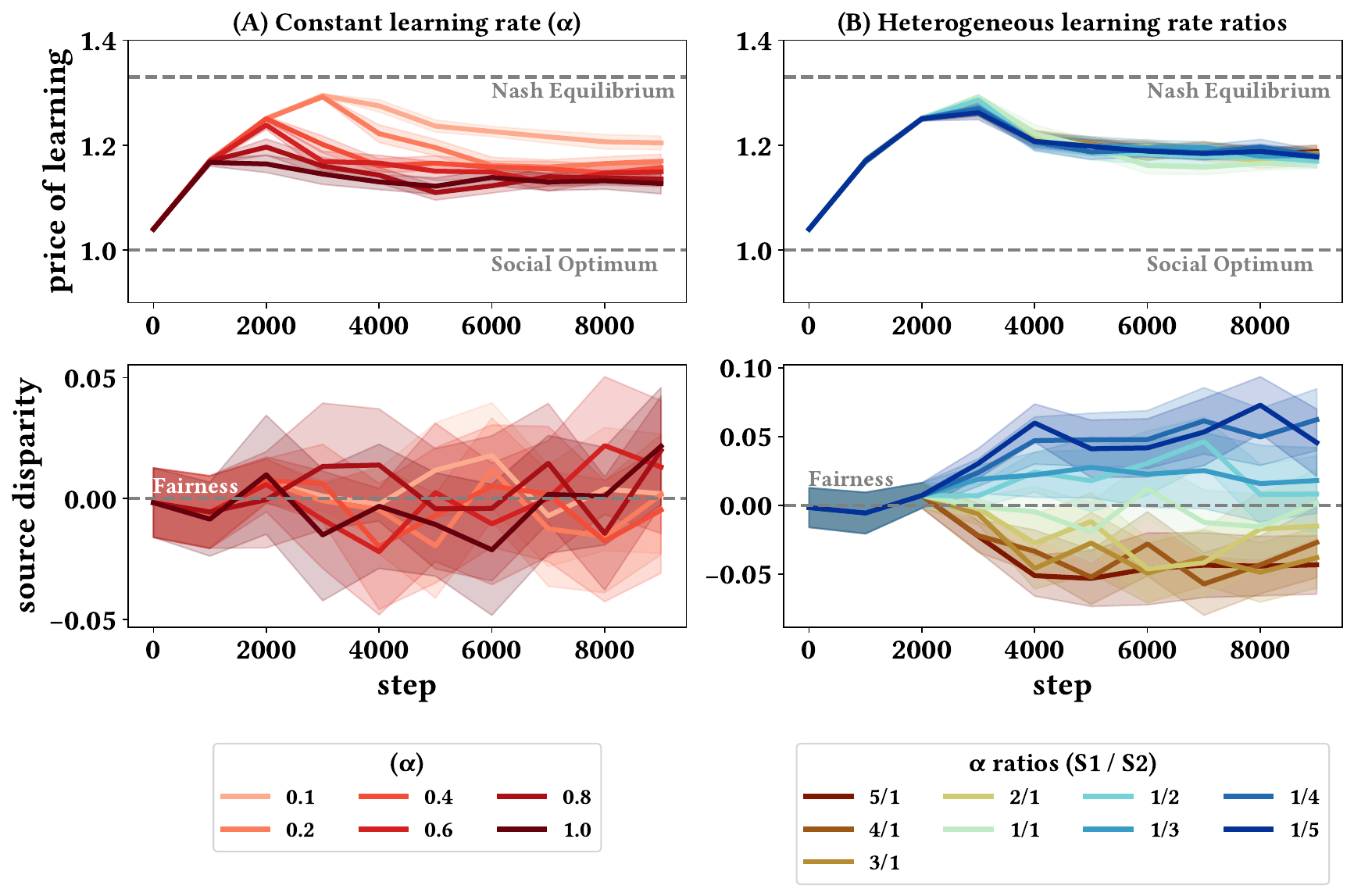}
    \caption{\textbf{Top:} With equal learning rates, the system converges to a profile between the Social Optimum and the Nash Equilibrium. Source disparity fluctuates around zero, indicating fairness across sources. \textbf{Bottom:} With unequal rates, agents with higher learning rates gain a persistent advantage. The system converges to inequitable states even though the price of learning is comparable.}
    \label{fig:chapter5:fb_learning_rates}
    \vspace{-4mm}
\end{figure}

When agents have equal learning rates (\Cref{fig:chapter5:fb_learning_rates}, top), the system converges to a state whose total cost lies between the SO and the NE. However, source disparity oscillates around zero, indicating that agents from different sources experience similar travel times. In other words, equal learning rates prevent persistent inequality.

In contrast, when learning rates differ between sources (\Cref{fig:chapter5:fb_learning_rates}, bottom), disparities emerge. Agents with higher learning rates achieve lower travel times than slower-learning counterparts. This advantage persists throughout training and converges to a state of nonzero source disparity. Notably, this occurs even though the system’s overall price of learning remains similar, demonstrating that looking at aggregate efficiency metrics alone obscures underlying inequities.

These results highlight a crucial insight: efficiency metrics are insufficient to capture the distributional consequences of learning. Heterogeneous learning rates systematically favor some agents, producing inequitable outcomes even when overall efficiency appears unaffected. This contrasts sharply with equilibrium-based expectations and underscores the importance of explicitly accounting for adaptation heterogeneity in multi-agent learning systems.

\begin{figure}[h]
\centering
\includegraphics[width=4in]{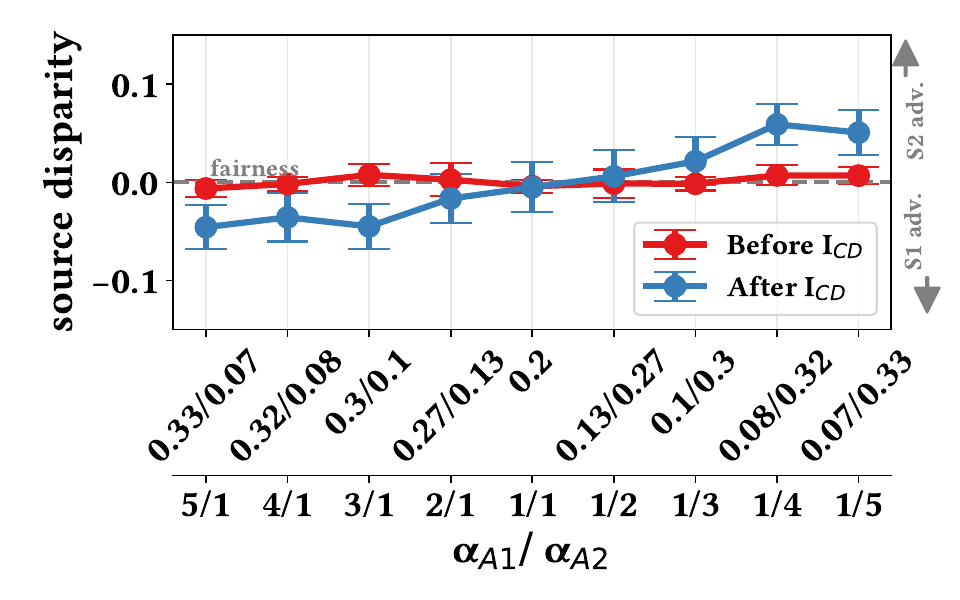}
    \caption{When varying the relative learning rates between the two sources, disparities emerge after the intervention: agents with slower learning rate experience worse travel times than the faster-learning ones. This effect is most pronounced when learning rates differ sharply (e.g., 5:1 or 1:5).}
    \label{fig:chapter5:learning_rates_ratios_0.2}
    \vspace{-4mm}
\end{figure}

\subsubsection{Network interventions exacerbate disparities under unequal learning rates}

\begin{figure*}[h!]
    \centerline{\includegraphics[width=7in]{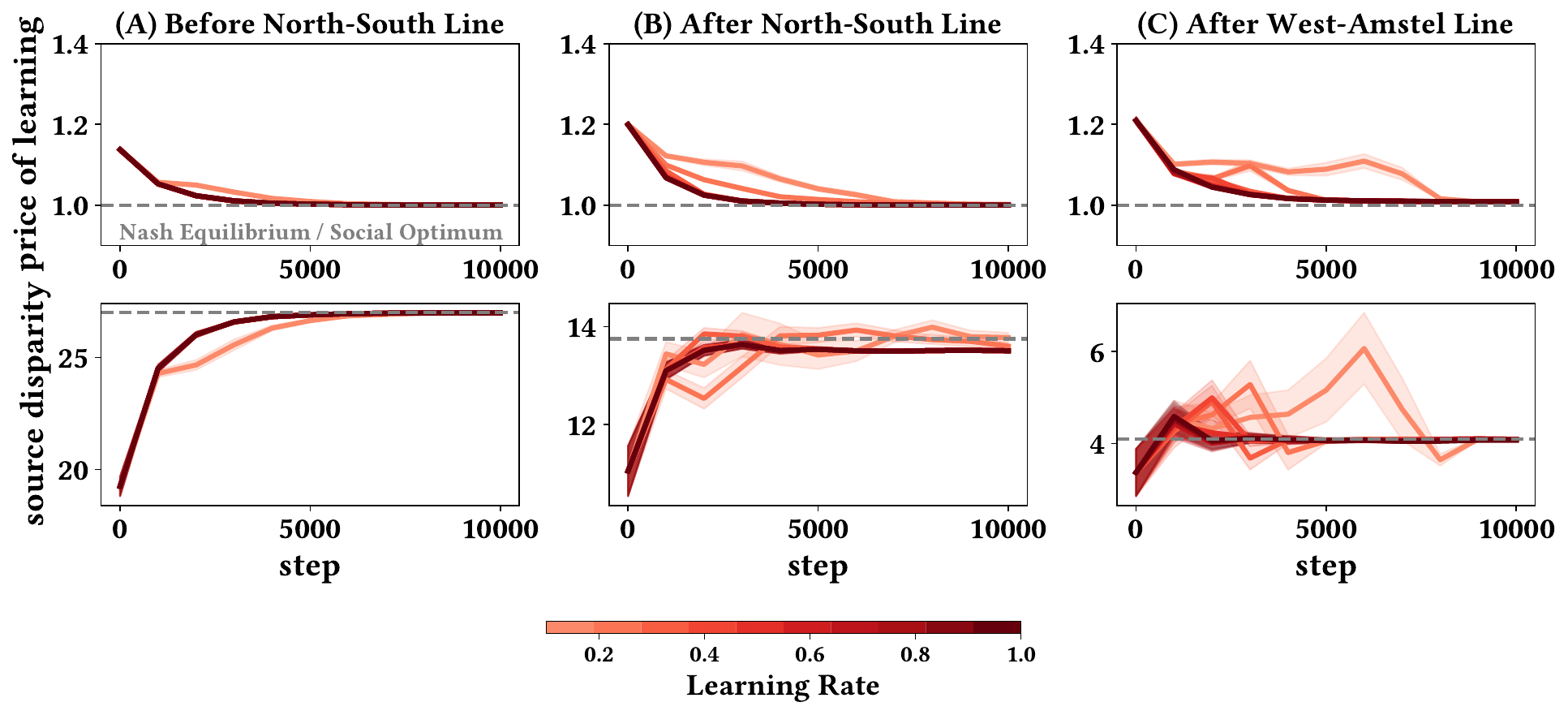}}
    \caption{Training under uniform learning rates  in the Amsterdam Metro network. We show the evolution of social cost and source disparity over training, for Phases A, B, and C. While the system eventually converges to the Nash Eq., learning inefficiencies and oscillations in source disparity highlight the unequal effects of adaptation even under homogeneous learning.}
    \label{fig:chapter5:amsterdam_all_fixed_learning_rates}
\end{figure*}

\Cref{fig:chapter5:learning_rates_ratios_0.2} shows how source disparity evolves before and after the fast intervention. To isolate the effects of heterogeneous learning rates, we fixed the average population learning rate at $\alpha = 0.2$ and varied the relative rates between sources. The intervention introduces disadvantages against the slower-learning agents, resulting in higher average travel times compared to faster-learning agents. This disparity intensifies when learning rates differ sharply (e.g., 5:1 or 1:5), producing pronounced inequities.

We now investigate whether these dynamics persist in the realistic Amsterdam Metro environment.

\subsection{Results on the Amsterdam Metro Network}

\subsubsection{Convergence under uniform learning rates exhibits learning inefficiencies but converges to the NE}
When all agents employ the same learning rate, the system converges to the Nash equilibrium (NE) in the long run (Fig.~\ref{fig:chapter5:amsterdam_all_fixed_learning_rates}). The price of learning (PoL) settles at the optimal value of $1.0$, confirming that homogeneous adaptation ensures eventual efficiency.

Nevertheless, during early and intermediate stages agents deviate from the NE, causing pronounced inefficiency. In the baseline configuration (Phase A, “Before North‑South line”) the PoL climbs well above $1.0$ before descending back toward unity. This inefficiency reflects that, even with identical learning rates, the collective adjustment requires many iterations to converge.

After the North–South line is introduced (Phase B), PoL still converges to $1.0$, but source disparity behaves differently: rather than settling at the expected NE level, it oscillates around a target value. The oscillatory pattern suggests that the added infrastructure creates feedback loops in routing choices, causing the two sources to alternately over‑ and under‑react to each other’s adjustments. With low learning rates, convergence to the NE is not guaranteed.

Overall, the uniform‑learning‑rate scenario provides a useful baseline: while agents eventually convergence to the Nash equilibrium, the transient dynamics can be inefficient.

\begin{figure*}[ht]
    \centerline{\includegraphics[width=7in]{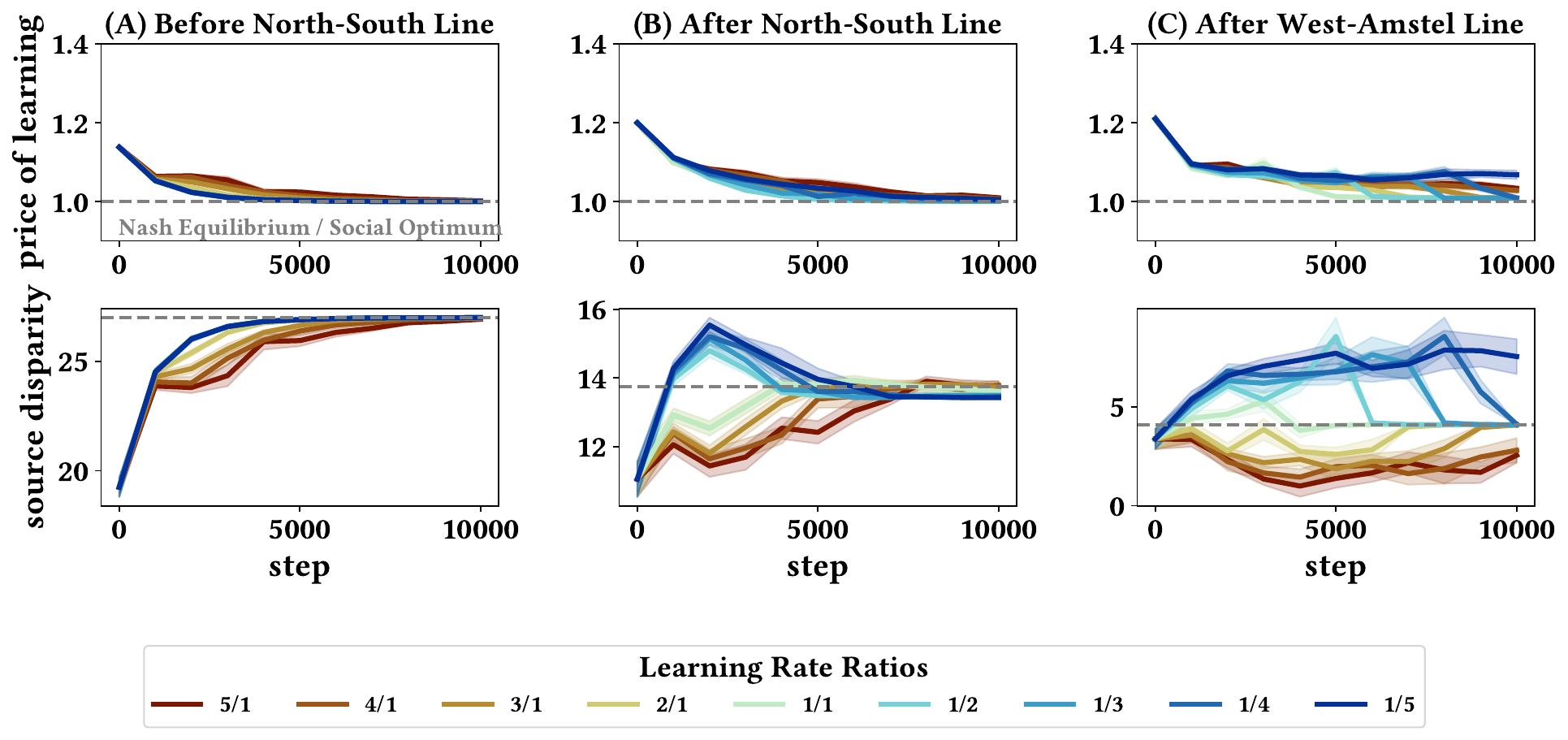}}
    \caption{Training under heterogeneous learning rates between West and East sources in the Amsterdam Metro network. Each plot shows the evolution of price of learning and source disparity over training steps for different learning rate ratios. Unequal learning rates produce persistent inefficiencies and systematic disparities, particularly after network expansions (Phase C), demonstrating that faster-adapting commuters gain a consistent advantage over slower learners.}
    \label{fig:chapter5:amsterdam_all_training_steps_split_rates}
\end{figure*}

\subsubsection{Unequal learning rates raise inefficiencies}
In Phase A and B, heterogeneous learning rates still allow eventual convergence to the Nash equilibrium: PoL converges to $1.0$ for all tested ratios. However, again, during a large portion of the learning process the system operates inefficiently. Under extreme imbalances (e.g., $5\!:\!1$ or $1\!:\!5$), PoL remains well above $1.0$ for thousands of steps before settling. Since commuters continuously adapt and seldom achieve optimal strategies instantly, these prolonged inefficiencies are practically significant.

Fairness exhibits a comparable time dependence. In Case A, disparities fade over time, yet faster learners retain a consistent advantage during the learning phase. In Case B, the impact is structural: when the East source learns faster (ratios $1/2$, $1/3$, $1/4$, $1/5$), the system converges to states where the West source endures higher average travel times than either the Nash or social optimum predicts. Thus, unequal learning not only prolongs inefficiency during the learning phase but also shifts fairness outcomes at convergence, producing persistent disadvantages for slower learners.

\subsubsection{Persistent inefficiencies and unfairness emerge after interventions}
The addition of the West–Amstel line changes the dynamics qualitatively. With this extension, unequal learning rates can block convergence to the Nash equilibrium, unlike Phases A and B where efficiency was eventually guaranteed. For some ratios, PoL stabilizes above $1.0$, indicating that persistent inefficiencies remain even after learning. Fairness is likewise impacted: source disparity may fail to converge to the levels predicted by Nash or socially optimal behaviour. In several instances, faster learners retain advantages, while slower learners suffer systematically higher travel times.

This differs from earlier phases: whereas learning inefficiencies eventually dissipated, in Case C they persist even after convergence, especially for high learning ratio differences. Infrastructure expansion therefore interacts with heterogeneous adaptation in unexpected ways, creating lasting inefficiencies and inequities, especially as the network gets larger.

\subsection{Key Takeaway}
Our results show that static Nash equilibrium analysis overlooks important adaptation dynamics. In both settings studied, unequal learning rates, likely to exist in reality, introduce persistent disparities between groups. Importantly, this disparities occur after network interventions. This highlights that a low Price of Anarchy does not guarantee equitable outcomes: faster-adapting commuters can systematically benefit from new routes, while slower learners face higher costs. For transport planning, this suggests that infrastructure expansions should be assessed not only for their equilibrium efficiency but also for adaptation dynamics. Planners can mitigate inequities by supporting slower-to-adapt groups---for example through targeted communication, real-time travel information, or phased rollouts---so that efficiency gains from new infrastructure are distributed more evenly.

\section{Conclusion}
We examined how heterogeneous learning dynamics shape efficiency and fairness in transport networks, using reinforcement-learning agents in congestion games. Our simulations revealed several key insights. First, differences in learning rates across commuter groups can produce persistent inequities: faster-learning agents can consistently experience lower travel times, even when overall system efficiency remains high. Second, network interventions interact with these heterogeneous behaviours in complex ways --- some expansions improve aggregate efficiency but disproportionately benefit faster learners, while others can reduce disparities at the cost of slower convergence or temporary inefficiencies.

This paper contributes to the ongoing research on fair transport network design \cite{martens2016transport,BEHBAHANI2019171,michailidis2023balancing}, by emphasizing the importance of considering sources of heterogeneity that dictate mobility behaviour. We highlight the interplay between individual learning and systemic fairness. Our results suggest that interventions aimed at mitigating congestion should also account for the heterogeneity in agents’ learning, and overlooking that reinforces inequities. 

Future work could extend our analysis to asymmetric networks where, even with fully rational agents, a trade-off exists between optimal and fair routing---as demonstrated in \cite{roughgarden2002unfair}. Furthermore,  it would be relevant to consider alternative learning algorithms, inspired in psychology and economics literature, such as  Roth-Erev learning \cite{erev1998predicting} or model-based algorithms \cite{daw2011model}. Investigating how such inequities might be mitigated in the presence of learning agents remains an important direction for research.

\section*{Acknowledgments}
This project was funded by the Innovation Center for Artificial Intelligence (ICAI) and the City of Amsterdam.

%Bibliography
\bibliographystyle{unsrt}  
\bibliography{references}

@ARTICLE{carissimo_counter_intuitive,
  author={Carissimo, Cesare},
  journal={IEEE Access}, 
  title={Counter-Intuitive Effects of Q-Learning Exploration in a Congestion Dilemma}, 
  year={2024},
  volume={12},
  number={},
  pages={15984-15996},
  doi={10.1109/ACCESS.2024.3358608}}

@article{roughgarden_severity_2006,
	title = {On the severity of {Braess}'s {Paradox}: {Designing} networks for selfish users is hard},
	volume = {72},
	issn = {0022-0000},
	url = {https://www.sciencedirect.com/science/article/pii/S0022000006000109},
	doi = {https://doi.org/10.1016/j.jcss.2005.05.009},
	number = {5},
	journal = {Journal of Computer and System Sciences},
	author = {Roughgarden, Tim},
	year = {2006},
	pages = {922--953},
}

@article{gairing_network_design_2017,
author = {Gairing, Martin and Harks, Tobias and Klimm, Max},
title = {Complexity and Approximation of the Continuous Network Design Problem},
journal = {SIAM Journal on Optimization},
volume = {27},
number = {3},
pages = {1554-1582},
year = {2017},
doi = {10.1137/15M1016461},

URL = { 
    
        https://doi.org/10.1137/15M1016461
    
    

},
eprint = { 
    
        https://doi.org/10.1137/15M1016461
    
    

}
}

@article{falleur_income_behavior,
author = {Boon-Falleur, Melusine and Baumard, Nicolas and André, Jean-Baptiste},
year = {2024},
month = {01},
pages = {17456916231201512},
title = {The Effect of Income and Wealth on Behavioral Strategies, Personality Traits, and Preferences},
journal = {Perspectives on psychological science : a journal of the Association for Psychological Science},
doi = {10.1177/17456916231201512}
}

@article{Zhang_coordinated_route_2023,
author = {Zhang, Le and Khalgui, Mohamed and Li, Zhiwu and Zheng, Shanshui and Zhang, Yongsheng},
title = {A unified model for the fairness mechanism-based coordinated vehicle route guidance},
journal = {IET Intelligent Transport Systems},
volume = {17},
number = {12},
pages = {2369-2380},
doi = {https://doi.org/10.1049/itr2.12414},
url = {https://ietresearch.onlinelibrary.wiley.com/doi/abs/10.1049/itr2.12414},
eprint = {https://ietresearch.onlinelibrary.wiley.com/doi/pdf/10.1049/itr2.12414},
year = {2023}
}

@article{milchtaich1996congestion,
  title={Congestion games with player-specific payoff functions},
  author={Milchtaich, Igal},
  journal={Games and economic behavior},
  volume={13},
  number={1},
  pages={111--124},
  year={1996},
  publisher={Elsevier}
}

@inproceedings{roughgarden2002unfair,
  title={How unfair is optimal routing?},
  author={Roughgarden, Tim},
  booktitle={Symposium on Discrete Algorithms: Proceedings of the thirteenth annual ACM-SIAM symposium on Discrete algorithms},
  volume={6},
  number={08},
  pages={203--204},
  year={2002}
}

@article{fischer2023fair,
  title={Fair Interventions in Weighted Congestion Games},
  author={Fischer, Miriam and Gairing, Martin and Paccagnan, Dario},
  journal={arXiv preprint arXiv:2311.16760},
  year={2023}
}

@inproceedings{ferguson2021impact,
  title={The impact of fairness on performance in congestion networks},
  author={Ferguson, Bryce L and Marden, Jason R},
  booktitle={2021 American control conference (ACC)},
  pages={4521--4526},
  year={2021},
  organization={IEEE}
}

@inproceedings{oesterle2024raise,
  title={RAISE the Bar: Restriction of Action Spaces for Improved Social Welfare and Equity in Traffic Management},
  author={Oesterle, Michael and Grams, Tim and Bartelt, Christian and Stuckenschmidt, Heiner},
  booktitle={Proceedings of the 23rd International Conference on Autonomous Agents and Multiagent Systems},
  pages={1492--1500},
  year={2024}
}

@article{Belov04072022,
author = {Aleksandr Belov and Konstantinos Mattas and Michail Makridis and Monica Menendez and Biagio Ciuffo},
title = {A microsimulation based analysis of the price of anarchy in traffic routing: The enhanced Braess network case},
journal = {Journal of Intelligent Transportation Systems},
volume = {26},
number = {4},
pages = {448--460},
year = {2022},
publisher = {Taylor \& Francis},
doi = {10.1080/15472450.2021.1904920},


URL = { 
    
        https://doi.org/10.1080/15472450.2021.1904920
    
    

},
eprint = { 
    
        https://doi.org/10.1080/15472450.2021.1904920
    
    

}

}

@article{braess_paradox,
 ISSN = {00411655, 15265447},
 URL = {http://www.jstor.org/stable/25769266},
 author = {Dietrich Braess and Anna Nagurney and Tina Wakolbinger},
 journal = {Transportation Science},
 number = {4},
 pages = {446--450},
 publisher = {INFORMS},
 title = {On a Paradox of Traffic Planning},
 urldate = {2025-01-19},
 volume = {39},
 year = {2005}
}

@book{sutton_reinforcement_2018,
	address = {Cambridge, Massachusetts},
	edition = {Second edition},
	series = {Adaptive computation and machine learning series},
	title = {Reinforcement learning: an introduction},
	isbn = {978-0-262-03924-6},
	shorttitle = {Reinforcement learning},
	language = {en},
	publisher = {The MIT Press},
	author = {Sutton, Richard S. and Barto, Andrew G.},
	year = {2018},
}

@article{monderer_potential_1996,
	title = {Potential {Games}},
	volume = {14},
	issn = {0899-8256},
	url = {https://www.sciencedirect.com/science/article/pii/S0899825696900445},
	doi = {https://doi.org/10.1006/game.1996.0044},
	number = {1},
	journal = {Games and Economic Behavior},
	author = {Monderer, Dov and Shapley, Lloyd S.},
	year = {1996},
	pages = {124--143},
}

@article{bonau_begavioural_game_theory,
author = {Bonau, Sarah},
year = {2017},
month = {02},
pages = {7},
title = {A Case for Behavioural Game Theory},
volume = {6},
doi = {10.5923/j.jgt.20170601.02}
}

@article{lee_game_2008,
	title = {Game theory and neural basis of social decision making},
	volume = {11},
	issn = {1097-6256},
	url = {https://www.ncbi.nlm.nih.gov/pmc/articles/PMC2413175/},
	doi = {10.1038/nn2065},
	number = {4},
	urldate = {2025-01-20},
	journal = {Nature neuroscience},
	author = {Lee, Daeyeol},
	month = apr,
	year = {2008},
	pmid = {18368047},
	pmcid = {PMC2413175},
	pages = {404--409},
}

@article{selten_commuters_2007,
	title = {Commuters route choice behaviour},
	volume = {58},
	issn = {0899-8256},
	url = {https://www.sciencedirect.com/science/article/pii/S089982560600039X},
	doi = {https://doi.org/10.1016/j.geb.2006.03.012},
	number = {2},
	journal = {Games and Economic Behavior},
	author = {Selten, R. and Chmura, T. and Pitz, T. and Kube, S. and Schreckenberg, M.},
	year = {2007},
	pages = {394--406},
}

@article{chremos_traveler-centric_2023,
	title = {A traveler-centric mobility game: {Efficiency} and stability under rationality and prospect theory},
	volume = {18},
	url = {https://doi.org/10.1371/journal.pone.0285322},
	doi = {10.1371/journal.pone.0285322},
	number = {5},
	journal = {PLOS ONE},
	author = {Chremos, Ioannis Vasileios and Malikopoulos, Andreas A.},
	month = may,
	year = {2023},
	note = {Publisher: Public Library of Science},
	pages = {1--32},
}

@article{chremos2022design,
  title={The design and analysis of a mobility game},
  author={Chremos, Ioannis Vasileios and Malikopoulos, Andreas A},
  journal={arXiv preprint arXiv:2202.07691},
  year={2022}
}

@inproceedings{roman2019multi,
  title={Multi-population congestion games with incomplete information},
  author={Roman, Charlotte and Turrini, Paolo},
  booktitle={Proceedings of the Twenty-Eighth International Joint Conference on Artificial Intelligence},
  pages={565--571},
  year={2019},
  organization={International Joint Conferences on Artificial Intelligence}
}

@article{cats2020learning,
  title={Learning and adaptation in dynamic transit assignment models for congested networks},
  author={Cats, Oded and West, Jens},
  journal={Transportation Research Record},
  volume={2674},
  number={1},
  pages={113--124},
  year={2020},
  publisher={SAGE Publications Sage CA: Los Angeles, CA}
}

@article{ben-elia_which_2010,
	title = {Which road do {I} take? {A} learning-based model of route-choice behavior with real-time information},
	volume = {44},
	issn = {0965-8564},
	url = {https://www.sciencedirect.com/science/article/pii/S0965856410000170},
	doi = {https://doi.org/10.1016/j.tra.2010.01.007},
	number = {4},
	journal = {Transportation Research Part A: Policy and Practice},
	author = {Ben-Elia, Eran and Shiftan, Yoram},
	year = {2010},
	pages = {249--264},
}

@article{chotibut2020route,
  title={The route to chaos in routing games: When is price of anarchy too optimistic?},
  author={Chotibut, Thiparat and Falniowski, Fryderyk and Misiurewicz, Micha{\l} and Piliouras, Georgios},
  journal={Advances in Neural Information Processing Systems},
  volume={33},
  pages={766--777},
  year={2020}
}

@inproceedings{bielawski2021follow,
  title={Follow-the-regularized-leader routes to chaos in routing games},
  author={Bielawski, Jakub and Chotibut, Thiparat and Falniowski, Fryderyk and Kosiorowski, Grzegorz and Misiurewicz, Micha{\l} and Piliouras, Georgios},
  booktitle={International Conference on Machine Learning},
  pages={925--935},
  year={2021},
  organization={PMLR}
}

@inproceedings{wunder2010classes,
  title={Classes of multiagent q-learning dynamics with epsilon-greedy exploration},
  author={Wunder, Michael and Littman, Michael L and Babes, Monica},
  booktitle={Proceedings of the 27th International Conference on Machine Learning (ICML-10)},
  pages={1167--1174},
  year={2010}
}

@inproceedings{ramos2020toll,
  title={Toll-based learning for minimising congestion under heterogeneous preferences},
  author={Ramos, Gabriel de O and R{\u{a}}dulescu, Roxana and Now{\'e}, Ann and Tavares, Anderson R},
  booktitle={Proceedings of the 19th International Conference on Autonomous Agents and MultiAgent Systems},
  pages={1098--1106},
  year={2020}
}

@article{ahmad2024travel,
  title={Travel behaviour and game theory: A review of route choice modeling behaviour},
  author={Ahmad, Furkan and Al-Fagih, Luluwah},
  journal={Journal of choice modelling},
  volume={50},
  pages={100472},
  year={2024},
  publisher={Elsevier}
}

@article{wei_day_to_day,
author = {Wei, Fangfang and Ma, Shoufeng and Jia, Ning},
title = {A Day-to-Day Route Choice Model Based on Reinforcement Learning},
journal = {Mathematical Problems in Engineering},
volume = {2014},
number = {1},
pages = {646548},
doi = {https://doi.org/10.1155/2014/646548},
url = {https://onlinelibrary.wiley.com/doi/abs/10.1155/2014/646548},
eprint = {https://onlinelibrary.wiley.com/doi/pdf/10.1155/2014/646548},
year = {2014}
}

@inproceedings{de_o_ramos_improved_2015,
	address = {Berlin, Heidelberg},
	title = {An {Improved} {Learning} {Automata} {Approach} for the {Route} {Choice} {Problem}},
	isbn = {978-3-662-46241-6},
	booktitle = {Agent {Technology} for {Intelligent} {Mobile} {Services} and {Smart} {Societies}},
	publisher = {Springer Berlin Heidelberg},
	author = {de O. Ramos, Gabriel and Grunitzki, Ricardo},
	editor = {Koch, Fernando and Meneguzzi, Felipe and Lakkaraju, Kiran},
	year = {2015},
	pages = {56--67},
}

@article{levy2018emergence,
  title={Emergence of cooperation and a fair system optimum in road networks: A game-theoretic and agent-based modelling approach},
  author={Levy, Nadav and Klein, Ido and Ben-Elia, Eran},
  journal={Research in Transportation Economics},
  volume={68},
  pages={46--55},
  year={2018},
  publisher={Elsevier}
}

@book{martens2016transport,
  title={Transport justice: Designing fair transportation systems},
  author={Martens, Karel},
  year={2016},
  publisher={Routledge}
}

@article{baker2009removing,
  title={Removing roads and traffic lights speeds urban travel},
  author={Baker, Linda},
  journal={Scientific American},
  volume={1},
  year={2009}
}

@article{pedroso2024fair,
  title={Fair artificial currency incentives in repeated weighted congestion games: Equity vs. equality},
  author={Pedroso, Leonardo and Agazzi, Andrea and Heemels, WPMH and Salazar, Mauro},
  journal={arXiv preprint arXiv:2403.03999},
  year={2024}
}

@inproceedings{censi2019today,
  title={Today me, tomorrow thee: Efficient resource allocation in competitive settings using karma games},
  author={Censi, Andrea and Bolognani, Saverio and Zilly, Julian G and Mousavi, Shima Sadat and Frazzoli, Emilio},
  booktitle={2019 IEEE Intelligent Transportation Systems Conference (ITSC)},
  pages={686--693},
  year={2019},
  organization={IEEE}
}

@article{dong2020segregated,
  title={Segregated interactions in urban and online space},
  author={Dong, Xiaowen and Morales, Alfredo J and Jahani, Eaman and Moro, Esteban and Lepri, Bruno and Bozkaya, Burcin and Sarraute, Carlos and Bar-Yam, Yaneer and Pentland, Alex},
  journal={EPJ Data Science},
  volume={9},
  number={1},
  pages={20},
  year={2020},
  publisher={Springer Berlin Heidelberg}
}

@article{youn2008price,
  title={Price of anarchy in transportation networks: efficiency and optimality control},
  author={Youn, Hyejin and Gastner, Michael T and Jeong, Hawoong},
  journal={Physical review letters},
  volume={101},
  number={12},
  pages={128701},
  year={2008},
  publisher={APS}
}

@article{castelnovo_clarification_2022,
	title = {A clarification of the nuances in the fairness metrics landscape},
	volume = {12},
	issn = {2045-2322},
	url = {https://doi.org/10.1038/s41598-022-07939-1},
	doi = {10.1038/s41598-022-07939-1},
	number = {1},
	journal = {Scientific Reports},
	author = {Castelnovo, Alessandro and Crupi, Riccardo and Greco, Greta and Regoli, Daniele and Penco, Ilaria Giuseppina and Cosentini, Andrea Claudio},
	month = mar,
	year = {2022},
	pages = {4209},
}

@article{SHOHAM2007365,
title = {If multi-agent learning is the answer, what is the question?},
journal = {Artificial Intelligence},
volume = {171},
number = {7},
pages = {365-377},
year = {2007},
note = {Foundations of Multi-Agent Learning},
issn = {0004-3702},
doi = {https://doi.org/10.1016/j.artint.2006.02.006},
url = {https://www.sciencedirect.com/science/article/pii/S0004370207000495},
author = {Yoav Shoham and Rob Powers and Trond Grenager}
}

@article{erev1998predicting,
  title={Predicting how people play games: Reinforcement learning in experimental games with unique, mixed strategy equilibria},
  author={Erev, Ido and Roth, Alvin E},
  journal={American Economic Review},
  pages={848--881},
  year={1998}
}

@article{daw2011model,
  title={Model-based influences on humans' choices and striatal prediction errors},
  author={Daw, Nathaniel D and Gershman, Samuel J and Seymour, Ben and Dayan, Peter and Dolan, Raymond J},
  journal={Neuron},
  volume={69},
  number={6},
  pages={1204--1215},
  year={2011},
  publisher={Elsevier}
}

@article{steinberg1983prevalence,
  title={The prevalence of Braess' paradox},
  author={Steinberg, Richard and Zangwill, Willard I},
  journal={Transportation Science},
  volume={17},
  number={3},
  pages={301--318},
  year={1983},
  publisher={INFORMS}
}

@article{claus1998dynamics,
  title={The dynamics of reinforcement learning in cooperative multiagent systems},
  author={Claus, Caroline and Boutilier, Craig},
  journal={AAAI/IAAI},
  volume={1998},
  number={746-752},
  pages={2},
  year={1998}
}

@article{BEHBAHANI2019171,
title = {A conceptual framework to formulate transportation network design problem considering social equity criteria},
journal = {Transportation Research Part A: Policy and Practice},
volume = {125},
pages = {171-183},
year = {2019},
issn = {0965-8564},
doi = {https://doi.org/10.1016/j.tra.2018.04.005},
url = {https://www.sciencedirect.com/science/article/pii/S0965856417308030},
author = {Hamid Behbahani and Sobhan Nazari and Masood {Jafari Kang} and Todd Litman}
}

@inproceedings{michailidis2023balancing,
  title={Balancing fairness and efficiency in transport network design through reinforcement learning},
  author={Michailidis, Dimitris and Ghebreab, Sennay and Santos, Fernando P},
  booktitle={Proceedings of the 2023 International Conference on Autonomous Agents and Multiagent Systems},
  pages={2532--2534},
  year={2023}
}

@article{afsar_rl_survey,
author = {Afsar, M. Mehdi and Crump, Trafford and Far, Behrouz},
title = {Reinforcement Learning based Recommender Systems: A Survey},
year = {2022},
issue_date = {July 2023},
publisher = {Association for Computing Machinery},
address = {New York, NY, USA},
volume = {55},
number = {7},
issn = {0360-0300},
url = {https://doi.org/10.1145/3543846},
doi = {10.1145/3543846},
journal = {ACM Comput. Surv.},
month = dec,
articleno = {145},
numpages = {38}
}

@article{li_crowding_pt,
title = {Crowding in Public Transport: A Review of Objective and Subjective Measures},
journal = {Journal of Public Transportation},
volume = {16},
number = {2},
pages = {107-134},
year = {2013},
issn = {1077-291X},
doi = {https://doi.org/10.5038/2375-0901.16.2.6},
url = {https://www.sciencedirect.com/science/article/pii/S1077291X22012486},
author = {Zheng Li and David A. Hensher}
}

@article{das_link_cost_function,
author = {Aathira K. Das and Bhargava Rama Chilukuri},
title ={Link Cost Function and Link Capacity for Mixed Traffic Networks},
journal = {Transportation Research Record},
volume = {2674},
number = {9},
pages = {38-50},
year = {2020},
doi = {10.1177/0361198120926454},
URL = { 
        https://doi.org/10.1177/0361198120926454
},
eprint = { 
        https://doi.org/10.1177/0361198120926454
}
}

@techreport{monitor_sociale_2023,
  author       = {{Redactie O\&S}},
  title        = {Monitor Sociale Veiligheid Openbaar Vervoer: Tussenmeting 2023},
  institution  = {Gemeente Amsterdam, Onderzoek \& Statistiek (O\&S)},
  year         = {2023},
  month        = {november},
  note         = {Behandeld in Commissie Algemene Zaken 5 december 2024},
  url          = {https://openresearch.amsterdam/nl/page/115463/monitor-sociale-veiligheid-openbaar-vervoer-tussenmeting-2023}
}

@online{gvb_north_south_line,
  author    = {{GVB}},
  title     = {De Noord/Zuidlijn is (bijna) jarig!},
  year      = {2019},
  url       = {https://over.gvb.nl/nieuws/de-noord-zuidlijn-is-jarig/},
  note      = {Geraadpleegd op 17 september 2025}
}

@article{rosenthal1973network,
  title={The network equilibrium problem in integers},
  author={Rosenthal, Robert W},
  journal={Networks},
  volume={3},
  number={1},
  pages={53--59},
  year={1973},
  publisher={Wiley Online Library}
}

@article{roughgarden_2002_how_bad_selfish_routing,
author = {Roughgarden, Tim and Tardos, \'{E}va},
title = {How bad is selfish routing?},
year = {2002},
issue_date = {March 2002},
publisher = {Association for Computing Machinery},
address = {New York, NY, USA},
volume = {49},
number = {2},
issn = {0004-5411},
url = {https://doi.org/10.1145/506147.506153},
doi = {10.1145/506147.506153},
journal = {J. ACM},
month = mar,
pages = {236–259},
numpages = {24}
}

\end{document}